\documentclass[12pt, letterpaper]{article}
\usepackage[top=1in, bottom=1.5in, left=1in, right=1in]{geometry}
\usepackage[utf8]{inputenc}
\usepackage{aas_macros}
\usepackage{booktabs}
\usepackage{hyperref}
\hypersetup{
    colorlinks=true,
    linkcolor=blue,
    citecolor=blue,
    allcolors=blue
}
\usepackage{color}
\usepackage{graphicx}
\usepackage{caption}
\usepackage{lineno}
\usepackage{comment}

\newcommand{\review}[1]{{\textcolor{black}{#1}}}

\title{Optimizing the LSST Observing Strategy for Dark Energy Science: DESC Recommendations for the Wide-Fast-Deep Survey}

\author{
\parbox{0.9\textwidth}{
\normalsize
Michelle~Lochner$^{1,2}$,
Daniel~M.~Scolnic$^{3}$,
Humna Awan$^{4}$,
Nicolas Regnault$^{5}$,
Philippe~Gris$^{6}$,
Rachel~Mandelbaum$^{7}$,
Eric~Gawiser$^{4}$,
Husni~Almoubayyed$^{7}$,
Christian~N.~Setzer$^{8,9}$,
Simon~Huber$^{10,11}$,
Melissa~L.~Graham$^{12}$,
Ren\'ee~Hlo\v{z}ek$^{13,14}$,
Rahul~Biswas$^{9}$,
Tim Eifler$^{15}$,
Daniel Rothchild$^{16}$,
Tarek~Allam Jr$^{17}$,
Jonathan~Blazek$^{18,19}$,
Chihway Chang$^{20}$,
Thomas Collett$^{21}$,
Ariel Goobar$^{9}$,
Isobel~M.~Hook$^{22}$,
Mike~Jarvis$^{23}$,
Saurabh~W.~Jha$^{4}$,
Alex~G.~Kim$^{24}$,
Phil~Marshall$^{25}$,
Jason~D.~McEwen$^{17}$,
Marc~Moniez$^{26}$,
Jeffrey~A.~Newman$^{27}$,
Hiranya~V.~Peiris$^{9,28}$,
Tanja~Petrushevska$^{29}$,
Jason~Rhodes$^{30}$,
Ignacio~Sevilla-Noarbe$^{31}$,
An\v{z}e~Slosar$^{32}$,
Sherry~H.~Suyu$^{10,11,33}$,
J.~Anthony~Tyson$^{34}$,
Peter~Yoachim$^{35}$
\begin{center}(The LSST Dark Energy Collaboration)\end{center}
}}
\date{}

\begin{document}

\maketitle

\begin{abstract}
Cosmology is one of the four science pillars of LSST, which promises to be transformative for our understanding of dark energy and dark matter. The LSST Dark Energy Science Collaboration (DESC) has been tasked with deriving constraints on cosmological parameters from LSST data. Each of the cosmological probes for LSST is heavily impacted by the choice of observing strategy.
This white paper is written by the LSST DESC Observing Strategy Task Force (OSTF), which represents the entire collaboration, and aims to make recommendations on observing strategy that will benefit all cosmological analyses with LSST. It is accompanied by the DESC DDF (Deep Drilling Fields) white paper (Scolnic et al.). We use a variety of metrics to understand the effects of the observing strategy on measurements of weak lensing, large-scale structure, clusters, photometric redshifts, supernovae, strong lensing and kilonovae. In order to reduce systematic uncertainties, we conclude that the current baseline observing strategy needs to be significantly modified to result in the best possible cosmological constraints.  
We provide some key recommendations: moving the WFD (Wide-Fast-Deep) footprint to avoid regions of high extinction, taking visit pairs in different filters, changing the 2$\times$15s snaps to a single exposure to improve efficiency, focusing on strategies that reduce long gaps ($>$15 days) between observations, and prioritizing spatial uniformity at several intervals during the 10-year survey.
\end{abstract}

\newpage
\section{White Paper Information}
Michelle Lochner, \url{dr.michelle.lochner@gmail.com}\\
Dan Scolnic, \url{dscolnic@kicp.uchicago.edu}\\
On behalf of the LSST Dark Energy Science Collaboration (DESC)
\\

\begin{enumerate} 
\item {\bf Science Category:}
\begin{itemize}
\item{The Nature of Dark Matter and Understanding Dark Energy} 
\item{Exploring the Changing Sky} 
\end{itemize}
\item {\bf Survey Type Category:}
    \begin{itemize}
    \item Wide-fast-deep
    \end{itemize}
\item {\bf Observing Strategy Category:}
\begin{itemize}
\item An integrated program with science that hinges on the combination of pointing and detailed 
	observing strategy - we propose a set of factors crucial for an observing strategy optimized for cosmology.
	\end{itemize}
\end{enumerate}

\clearpage

\section{Scientific Motivation}


\subsection{Introduction}
Providing cutting-edge constraints on dark matter and dark energy models is one of the key science goals of LSST. The ability to take almost all the observations required with the same instrument, as well as having a large enough dataset to subdivide repeatedly, will minimize the systematic effects that can dominate cosmological constraints. \autoref{fig:contours} shows the expected constraints almost entirely from LSST data alone, highlighting the complementarity of the static science and transient science probes required to break degeneracies in cosmological constraints. If observing strategy is not optimized for both sets of probes, cosmology with LSST will be limited by these degeneracies. In addition to being world leading in the cosmological probes of weak lensing, large-scale structure, clusters and supernovae, LSST also will allow completely novel studies such as statistical tests of cosmological isotropy, and the use of rare objects like kilonovae and strongly lensed supernovae for independent cosmological constraints. It will provide an enduring legacy dataset of galaxies and transient events that will be studied for decades. However, all this will only be possible if the observing strategy of LSST is carefully optimized. Here we briefly summarize the main cosmological probes of LSST, separated into static science and transient science, as the probes within these two categories have broadly similar observing strategy requirements. Of special consideration are photometric redshifts (photo-$z$), which are critical in enabling all cosmological probes, since spectroscopic follow-up for all LSST galaxies and transients will be impossible. Thus we consider in this paper the effect of observing strategy on photo-$z$ measurements, which rely on obtaining sufficient depth in the $ugriz$ filter set across the footprint.

\subsection{Static Science}

\hspace{\parindent}{\bf Weak Lensing:} 
The sample of billions of galaxies produced by LSST will be by far the largest ever compiled \cite{lsst2009}. Measuring the cosmological weak gravitational lensing signal of this galaxy sample will yield unprecedented constraints on cosmology, as this is one of the most powerful and direct probes of cosmology. Systematic effects such as PSF modeling errors, photo-$z$ errors and photometric calibration will be of critical importance and, with its high image quality and large data volumes, LSST is  well placed for understanding and mitigating these effects. In addition to having similar depth and area requirements of large-scale structure measurements, weak lensing also requires a high number of visits to mitigate PSF model shape errors and other systematic effects.\\

\hspace{\parindent}{\bf Large-Scale Structure and Galaxy Clusters:} 
 The same enormous dataset of galaxies will allow the use of cutting-edge analysis techniques, such as galaxy-galaxy correlations, cross-correlation with weak lensing, counts of galaxy clusters, and cross correlation with the CMB \cite{lsst2009}. Large-scale structure studies will place constraints on primordial fluctuations that are competitive with those from the CMB and will allow stringent tests of non-Gaussianity. Achieving these science goals will require surveying a large area with low extinction and uniform depth to mitigate systematic errors. The number of clusters within a given mass range is also a powerful probe of the growth of structure and dark energy. Cluster counts are included in our analysis and have similar requirements to weak lensing and large-scale structure. \\

\subsection{Transient Science}
\hspace{\parindent}{\bf Supernovae:} 
LSST will produce a catalog of hundreds of thousands of type Ia supernovae (SNIa), the exact number being highly dependent on the observing strategy, allowing unprecedented tests of cosmology with supernovae alone \cite{lsst2009}.
The sheer number of objects will also enable the study of the evolution of supernova intrinsic properties with redshift, as well as the impact of galaxy type and environment on cosmological constraints. However, LSST supernova cosmology demands well-measured, high-cadence light curves to classify objects as type Ia and obtain accurate distances. Thus, the WFD supernova survey is one of the cosmological probes most sensitive to choice of observing strategy, requiring regular cadence, frequent filter changes (for instance ensuring visit pairs in a night are in different filters) and a season length of about 5 months. To allow for the possible impact of systematic errors such as incorrect photometric classification and redshift uncertainty, we use a more stringent quality cut than was used in the LSST DESC Science Requirements Document (SRD) \cite{descsrd}. \autoref{fig:lightcurves} shows, as an illustrative example, the effect of cadence on light curve quality in the absence of spectroscopic redshift information.\\

{\bf Strong Lensing:} 
Strongly lensed objects are both faint and rare, making them difficult to detect in most surveys. The LSST WFD survey, however, will be able to detect hundreds of strongly lensed quasars and supernovae, allowing not only independent tests of cosmology, but also novel extragalactic studies and detailed dark matter analyses. Lensed quasars and supernovae allow a unique test of distance-duality, one of the fundamental relations in cosmology, as well as independent constraints on the much-contested value of $H_0$. Maximizing the number of strongly lensed supernovae detected would require ensuring every field is observed yearly and a high frequency of visits.\\

{\bf Gravitational Waves:} 
LSST will be the best facility of its time for the serendipitous detection of kilonovae, the electromagnetic (EM) counterpart to a gravitational wave (GW) event \cite{scolnic2017}. While we also support a small amount of target-of-opportunity time outlined in the DESC DDF white paper, serendipitous detections offer the possibility of discovering kilonovae below the LIGO-Virgo SNR threshold, paving the way for EM-triggered searches for sub-threshold signals in archival GW data. Such studies would provide a better understanding of the kilonova population, which may be critical for assessing systematics in the measurement of the Hubble constant from standard sirens. Combined with observations from gravitational wave detectors, kilonovae are a promising cosmological probe, complementary to other probes and with similar observational requirements to supernovae.

\newpage
\begin{center}

\begin{minipage}{0.99\columnwidth}
    \centering
    \includegraphics[width=0.6\columnwidth]{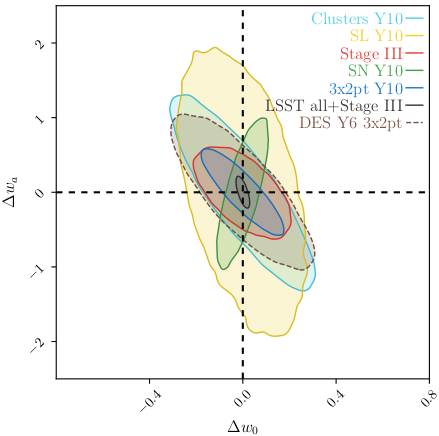}
    \captionof{figure}{Expected 10 year constraints on time-varying dark energy combining all LSST cosmological probes, from the DESC~SRD \cite{descsrd}. The contours labeled Stage III represent combined current constraints from the CMB, baryon acoustic oscillations, and supernovae. A possible realization of a DES Y6 3$\times$2pt analysis is included for comparison. LSST will clearly be a vast improvement over the current state of the art. }
    \label{fig:contours}
\end{minipage}

\vspace{20pt}

\begin{minipage}{0.99\columnwidth}
\centering
    \includegraphics[width=0.99\columnwidth]{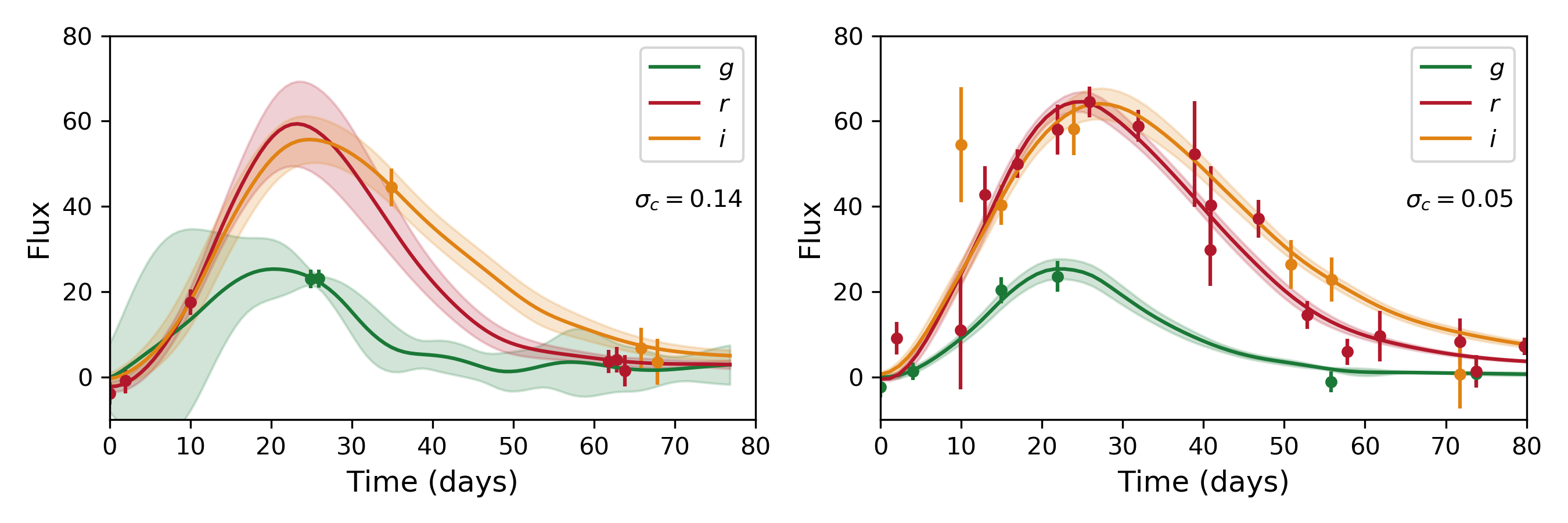}
    \captionof{figure}{An example SNIa light curve simulated with the \texttt{kraken\_2026} cadence (left) compared with a well-measured light curve with the proposed cadence in \autoref{sec:technical} (right). These are fit with a standard SALT2 \cite{guy2007} model without assuming a spectroscopic redshift. Also shown is the error on the color parameter, $\sigma_c$, which is a proxy for light curve quality.}
    \label{fig:lightcurves}
\end{minipage}

\end{center}

\vspace{.6in}

\section{Technical Description}
\label{sec:technical}

\subsection{High-level description}
In this section, we highlight several important changes to the observing strategy with LSST, also summarized in \autoref{tab:obs_constraints}, that would enable the best possible constraints on dark energy.
\begin{itemize}

    \item To improve all extragalactic science, we propose to move the 18,000 deg$^2$ WFD footprint away from the Galactic plane to avoid the zone of high extinction.  Galactic science has different observing strategy needs to extragalactic science, so we propose expanding the Galactic plane mini-survey and optimizing its cadence and filter choices for Milky Way science, including a microlensing survey to probe dark matter properties. We recommend a single exposure per visit instead of the 2$\times$15s snaps, which would result in approximately 7\% additional observing time, which could be used to expand the mini-surveys. Although the effect of exposure time on the cosmological probes is still being studied, 30s will likely be near the optimum.    
\item We stress that uniformity in co-added depth ($\sigma\sim0.1$, as defined in \autoref{sec:other_constraints}) across the entire WFD footprint is critical for cosmology and must be achieved after Y1, Y10 and at 2-3 reasonable intervals in between for data releases and intermediate analyses (here we use Years 1, 3, 6 and 10). 
\item To obtain light curves of sufficient quality for photometric classification of supernovae and for accurate distance measures, we find that a mean inter-night gap of 10($g$), 5($r$), 6($i$), 6($z$) days between observations is preferable. We require no long gaps ($>15$ days) within a season in $griz$. Enforcing this cadence results more than doubles the number of well-characterized\footnote{We use the following quality cuts: SNR$>5$ in at least 3 bands, 5 visits before peak and 10 after peak in any band and the error on the color parameter $\sigma_c<0.04$.} supernovae over the baseline strategy (see \autoref{fig:transients}).

\item Nightly visit pairs should be in neighboring filters instead of the same filter. 
\item To ensure a large number of complete light curves for supernovae and strongly lensed supernovae and quasars, longer season lengths\footnote{A season length is how long a field is observable in a year. It is largely dictated by the field's declination but can be quite short in observing strategies where low airmass is prioritized} are preferred, but not at the expense of the requested cadence. 
\item We are interested in exploring the possibility of redistributing roughly half of the $y$-band visits into $griz$. 
\item We note that a rolling cadence\footnote{Rolling cadence is a strategy where a part of the sky may be observed at higher cadence at the expense of another part of the sky. These areas are then exchanged after a period of time (usually one year).} may be required to achieve this frequency of inter-night visits but more study is needed to determine this. 
\item Finally, we find that results from AltSched, an alternative scheduler to OpSim, show that it is a highly promising approach and recommend exploring its scheduling methods in OpSim and especially comparing it to the new feature-based scheduler\footnote{The feature-based scheduler \cite{feature2018} uses a modified Markov decision process to choose observations on-demand, while the current OpSim proposal-based scheduler \cite{opsim2014} attempts to execute a predefined list of observations, similar to traditional manual telescope scheduling of proposals from various astronomers.} (see \autoref{sec:requests}).
\end{itemize}

\vspace{.3in}

\subsection{Footprint -- pointings, regions and/or constraints}
\label{sec:footprint}
All cosmological probes require observations at low Galactic extinction. Therefore, we strongly recommend selecting an 18,000 deg$^2$  footprint based on a E(B-V)$<0.2$ cut\footnote{In this analysis E(B-V) is derived from SFD maps \cite{sfd1998}  using MAF \cite{jones2014}.}. \autoref{fig:footprint} shows our proposed footprint in a declination range of $-70<$dec$<12.5$, chosen based on the extinction cut and to improve overlap with the 4-metre Multi-Object Spectroscopic Telescope (4MOST) \cite{4most2014} and the Dark Energy Spectroscopic Instrument (DESI) \cite{desi2014} surveys for spectroscopic follow-up. Our studies have shown that with the current WFD footprint, \emph{25\% of the area is unsuited for cosmology}. With our new proposed footprint, we will obtain considerable additional area for extragalactic science at no cost to depth\footnote{The strategy described in slide 6 of \url{http://ls.st/kak/} is close to our proposed footprint, although due to some regions with E(B-V)$>$0.2, it has less usable area for extragalactic science.}. \autoref{fig:static_fom} shows that the larger-area simulations such as \texttt{pontus\_2002} result in $\sim30\%$ more area which corresponds to a $\sim20\%$ improvement in dark energy constraints. However, the larger-area surveys have worse constraints on some systematics (see \autoref{sec:performance}) so we recommend to rather optimize the existing footprint than increase the total survey area. Determining the precise area that optimizes the trade-off between statistical and systematic errors is ongoing work. Critically, the proposed footprint increases overlap with the DESI survey from 3721 deg$^2$ to approximately 6000 deg$^2$, which will help ensure the DESC requirements on photo-$z$ calibration \cite{descsrd} are met using cross-correlations with DESI \cite{newman2013}. The footprint also maximizes overlap with the planned 4MOST TiDES survey, which will enable rapid spectroscopic follow-up of transients. The benefit of increased usable area and increased spectroscopic data will far outweigh the impact of increased atmospheric effects (including higher airmass) within the range of footprint changes simulated so far (which is anticipated to reduce co-added depth in the higher declination regions by only $\sim0.1$ mag).\\ 

\begin{minipage}{\columnwidth}
\centering
\includegraphics[width=0.8\columnwidth]{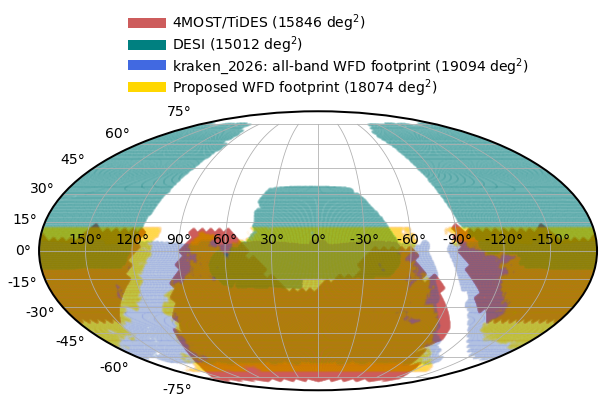}
 \captionof{figure}{Proposed WFD footprint based on an E(B-V)$<0.2$ cut (yellow), the baseline footprint (blue) and the DESI (green) and 4MOST/TiDES (red) footprints. The total area of each footprint is shown in parentheses.}  \label{fig:footprint}
\end{minipage}

\subsection{Image quality}
\label{sec:image_quality}
Weak lensing and other cosmological probes have stringent requirements on image quality, especially in $r$ and $i$. To produce the highest quality images, we recommend avoiding high airmass observations by scanning along the meridian as 
much 
as possible. However, this should generally not come at the cost of season length, which impacts transient science. Further potential optimizations to image quality include prioritizing certain filters in the best seeing conditions and varying exposure time based on observing conditions. These are described in \autoref{sec:requests}.

\subsection{Individual image depth and/or sky brightness}
A positive detection of supernovae requires a good single visit depth. The $5\sigma$ point source detection canonical baseline numbers from the LSST~SRD \cite{lsstSRD} should serve as reasonable goals ($g$:24.6, $r$:24.3, $i$:23.6, $z$:22.9).     

\subsection{Co-added image depth and/or total number of visits}
To achieve our goals with weak lensing, large-scale structure and clusters with the gold sample of galaxies as described in the Science Book \cite{lsst2009}, we require a co-added depth (for a $5\sigma$ point-source) of $i>24.5$ for Y1 and $i>26.0$ for Y10. 

Approaching 
sufficient co-added depth for $ugriz$ across the entire footprint is critical for photo-$z$, whilst  $y$  is somewhat less important (see \autoref{sec:filters}). The total number of visits is important to reduce weak lensing systematics (such as PSF shape modeling errors), which commonly
average down as a function of the number of visits. For transient-based cosmology, the cadence of visits across $griz$ is much more important than the total number. Thus we recommend as large a number of visits as possible, \emph{but only after all other requirements on co-added depth and cadence are met.}

\subsection{Number of visits within a night}
We strongly recommend visit pairs with different, neighboring filters. This improves the per-filter cadence and provides a color measurement that dramatically enhances transient classification, both of immense benefit to supernova cosmology. \emph{The loss of regular cadence due to multiple visits to the same field in a single night in the same filter would be highly damaging for cosmology.} We do \emph{not} advocate removing visit pairs (even though this improves some metrics; see \texttt{colossus\_2667} in \autoref{fig:transients}) so that asteroids can be identified separately from extragalactic transients and also to support Solar System science as one of the four pillars of LSST.

\subsection{Distribution of visits over time}
\label{sec:cadence}
To make optimal use of LSST for supernova cosmology, we recommend a mean inter-night gap of 10($g$), 5($r$), 6($i$), 6($z$) days. This recommendation is significantly higher cadence than \texttt{kraken\_2026} (22($g$), 12($r$), 12($i$), 14($z$)) but still less than the highest cadence strategy, \texttt{alt\_sched\_rolling} (8($g$), 3($r$), 4($i$), 3($z$)). Early classification of supernovae requires at least three points (the first having sufficient SNR to trigger an alert) in at least two bands during the rise of the light curve. We find that this is achievable with our recommended cadence for all observable SNe below $z=0.25$ (dropping off at $z\sim 0.4$) whereas \texttt{kraken\_2026} detects fewer than two points on the rise for more than 50\% of SNe, making early classification impossible. \\

However, average cadence is not the only consideration. Long gaps ($>$15 days) between observations in a season critically need to be avoided (see \autoref{fig:transients}, where the cadences are ordered from left to right with highest to lowest percentage of $r$-band visits separated by more than 15 days). The distribution of inter-night gaps must be narrow, and have a low average, to achieve excellent transient science results. Thus the filters should be cycled through as much as possible to create an even cadence. To accommodate rare, slowly varying transients such as strongly-lensed supernovae, longer season lengths (at least 5 months) are preferred as long as this cadence is achieved. This is shorter than the recommendations for the deep field, given that the WFD sample will probe wider areas of the sky to shallower redshifts.\\ 

To avoid negatively impacting static science, any rolling cadence must provide uniformity at the selected intervals (for example, Y1, Y3, Y6, Y10). \emph{A rolling cadence is thus only preferred if it is necessary to achieve the above cadence requirements.} While we do not currently have access to enough viable rolling cadence simulations to draw definitive conclusions, we remain very interested in continuing to investigate it as a potential strategy.

\subsection{Filter choice}
\label{sec:filters}
We are interested in investigating a change in the distribution of visits per filter. It is expected that a bluer distribution of filters would improve the number of high quality supernovae detected, without impacting static science. See \autoref{sec:requests} for a description of our proposed strategy. The filters $griz$ are all critically important for supernovae so should be regularly cycled through. $u$ band is very important for photo-$z$, while it may be less important to obtain the same depth in $y$ as the other bands \cite{graham2017}.

\subsection{Exposure constraints}
We strongly support dropping the 2$\times$15s snaps in favor of a single exposure as this increases observing efficiency while allowing a slower readout during slew, which will improve camera performance.  Rough calculations indicate the efficiency increases by about 7\%, which is significant given that (for example) the entire DDF program is only about 6\% of the total LSST observing time. This extra efficiency could be used to improve the cadence by increasing the number of inter-night visits or to expand the Galactic plane mini-survey to compensate for our proposed shifted footprint. Our analysis also indicates that it is unlikely that single 30s exposures will cause saturation for more than a small number of very nearby supernovae and so is still worth the efficiency gain. Additionally, the reduction (due to saturation) in the number of stars to be used for PSF modeling is not enough to have a significant impact on individual visit PSF modeling quality. While cosmic rays have been successfully removed in single exposures without snaps (e.g., with Hyper Suprime-Cam \cite{aihara2017}), we acknowledge that image simulations would be important to ensure the same can effectively be done with LSST.\\ 
Lastly, in our study, we found that \texttt{pontus\_2489} (an OpSim run with 20s exposures in $grizy$ and 40s in $u$) performed well for weak lensing systematics and did not decrease the depth and area. The sample of transients is larger but at a lower redshift on average. Further investigation would be needed to determine the optimal per-filter exposure time. 
We would also be interested in exploring variable exposures as described in \autoref{sec:requests}. 

\subsection{Other constraints}
\label{sec:other_constraints}
\begin{itemize}
\item {\bfseries Dithering: }Dithering is critical to delivering excellent 
calibration and 
cosmological constraints with LSST. We recommend random translational dithers of amplitude 0.5$\times$FOV for WFD with a single dither vector used on each night \cite{awan2016, carroll2014}.  We note that translational dithering can also be achieved with a no-fixed field strategy that re-tesselates the sky each night (such as the feature-based scheduler or AltSched). We also recommend that nightly rotational dithers be performed by choosing a single RotTelPos value randomly, returning to this value after each filter change and then controlling instrument rotation such that all re-visits to individual fields are performed at the same value of RotSkyPos. 
\item {\bfseries Uniformity: }We measure uniformity as the standard deviation of the $5\sigma$ $i$-band depth across the sky (after a depth cut). The Y10 uniformity of the baseline strategy \texttt{kraken\_2026} ($0.13$ mag) is sufficient, but we note that several strategies such as \texttt{mothra\_2045} are a factor of 2 less uniform, while AltSched results in a roughly 20\% improvement. Uniformity should be checked using this metric for key release years.
\item {\bfseries Synergy with deep drilling fields:} Achieving Stage IV dark energy goals will critically rely on the deep drilling fields (DDFs) of LSST. The DDFs are necessary for obtaining a sample of high-redshift, well-measured supernovae which improves cosmological constraints from SNIa by 36\% compared with WFD alone. High quality templates for training photo-$z$ algorithms and training samples for shear estimation both also require DDF observations. We refer the reader to the companion DESC DDF white paper (Scolnic et al. 2018) for a detailed description of the DDF requirements and note that the WFD and DDFs ultimately need to be optimized together.

\end{itemize}
\subsection{Estimated time requirement}
We request that the WFD survey utilize at least 80\% of total observing time
in order to achieve the LSST~SRD requirement of a median of 825 visits over at least 18,000 deg$^2$ along with our science goals.  As our new proposed WFD footprint would exclude the Galactic plane, we expect that the Galactic plane mini-survey will require increased time to achieve Milky Way science goals. 

\newpage

\begin{minipage}{\columnwidth}
    \centering
    \begin{tabular}{l|l|l|l}
        \toprule
        Properties & Importance \hspace{.3in} \\
        \midrule
        Image quality  &  1   \\
        \hspace{20pt} \emph{High quality especially in $r$ and $i$ and to a lesser extent $g$} & \\
        Sky brightness & 2 \\
        \hspace{20pt} \emph{Should not be prioritized over cadence} &\\
        Individual image depth &  2 \\
        \hspace{20pt} \emph{Sufficient to detect SNe} & \\
        Co-added image depth  &  1 \\
        \hspace{20pt} \emph{Uniformity is critical, depth as high as possible} & \\
        Number of exposures in a visit  & 2  \\
        \hspace{20pt} \emph{Should be 1 to improve efficiency} & \\
        Number of visits (in a night) & 1\\ 
        \hspace{20pt} \emph{Should be two visits per night in different filters} & \\ 
        Total number of visits &  2 \\
        \hspace{20pt} \emph{Should be as high as possible, after meeting other requirements}&\\
        Time between visits (in a night) & 2\\
        \hspace{20pt} \emph{Sufficient for asteroid detection ($\sim$40 minutes)} &  \\
        Time between visits (between nights)  &  1 \\
        \hspace{20pt} \emph{Should be around 10($g$),5($r$),6($i$),6($z$) days for SN characterization} & \\
        Long-term gaps between visits  & 1\\ 
        \hspace{20pt} \emph{$<15$ days within a season, season length of $\sim$5 months} &\\
        Dithering (translational and rotational) & 1\\
        \hspace{20pt} \emph{Critical to ensure uniformity} &\\
        Low extinction footprint & 1\\
        \hspace{20pt} \emph{Provides 25\% more area optimal for cosmology} &\\
        
        \bottomrule
    \end{tabular}
    \captionof{table}{{\bf Constraint rankings:} Summary of the relative importance of various survey strategy constraints (1=very important, 2=somewhat important, 3=not important). 
    }
        \label{tab:obs_constraints}
\end{minipage}

\subsection{Technical trades}
\begin{enumerate}
    \item 
    {\it What is the effect of a trade-off between your requested survey footprint (area) and requested co-added depth or number of visits?}\\
    \review{~~~~~The purely statistical constraining power of the joint static science probes has a well-described trade-off between depth and area, as can be seen in \autoref{fig:static_fom}. However, uniformity is still critical in reducing systematic effects and should not be reduced below that of the baseline observing strategy in order to gain extra area. Similarly, cosmology with supernovae only benefits from increased depth if it corresponds to improved cadence (for a given exposure time). Our proposed observing strategy in the above sections should allow excellent science with all cosmological probes.}
    \item {\it If not requesting a specific timing of visits, what is the effect of a trade-off between the uniformity of observations and the frequency of observations in time? e.g., a `rolling cadence' increases the frequency of visits during a short time period at the cost of fewer visits the rest of the time, making the overall sampling less uniform.}\\
    \review{~~~~~Uniformity is critical for the static science probes. However uniformity only needs to be ensured by specific release dates (for example, 1, 3, 6, 10 years) thus rolling cadence is still acceptable if it is done on the 2-4 year timescale. It should be noted that rolling cadence may not be a requirement if acceptable cadence can be achieved without it. The conclusion of the DESC is that rolling cadence remains an interesting observing strategy option that requires more in-depth simulation and study.}
    \item {\it What is the effect of a trade-off on the exposure time and number of visits (e.g., increasing the individual image depth but decreasing the overall number of visits)?}\\
    \review{~~~~~As long as sufficient image depth is achieved for supernovae and the exposure is not so short as to be read-out noise limited, more visits are preferred as it increases cadence and is helpful for weak lensing systematics. Removing the $2\times15$s snaps in favour of single exposure is a straightforward way to improve read-out noise and efficiency with few disadvantages.}
    \item {\it What is the effect of a trade-off between uniformity in number of visits and co-added depth? Is there any benefit to real-time exposure time optimization to obtain nearly constant single-visit limiting depth?}\\
    \review{~~~~~Uniformity in co-added depth is critical to static science and is more important than uniformity in the number of visits. While it does not matter if the number of visits is non-uniform across fields, this must not come at the cost of the cadence requirements described above. The option of real-time optimization may be beneficial,  
    but the effect must be investigated and there are likely 
    practical 
    limits to how much the exposure time can vary (due to read-out noise, image quality, saturation/non-linearity; see \autoref{sec:requests}).}
    \item {\it Are there any other potential trade-offs to consider when attempting to balance this proposal with others which may have similar but slightly different requests?}\\
    \review{~~~~~This proposal represents a coherent request from the entire DESC. However we recognize that the process of observing strategy optimization is ongoing. There is tension with Solar System science that drives the need for multiple nightly visits. However this would be mitigated for time domain cosmology if these visits were in different filters. We are also aware that our proposal for moving the footprint would be damaging to Milky Way science. This is mitigated by increasing the efficiency of the survey by replacing the 2$\times$15s snaps with single visits and using that time to expand the Galactic plane mini survey. Any different requests from other groups that result in OpSim runs can easily be run through our comprehensive set of metrics to examine the potential impact on cosmology.
    }
\end{enumerate}

\newpage
\section{Performance Evaluation}
\label{sec:performance}

We developed over 20 metrics to investigate the effects of the observing strategy on cosmology but due to space constraints, we only highlight a few of the most informative metrics. These will be fully described in an upcoming article (Lochner \emph{et al.} in prep). \\

{\bfseries The DETF Figure of Merit}:
The Dark Energy Task Force Figure of Merit (DETF FoM) is 
the reciprocal of the area of the contour enclosing 68\% of the credible interval constraining the dark energy parameters, $w_0$ and $w_a$, after marginalizing over other parameters \cite{albrecht2006}. We compute the FoM for the combined static probes, and for supernovae separately,
as described in \cite{descsrd}. Under conservative assumptions for photo-$z$ performance and/or host galaxy spectroscopic follow up, the dark energy FoM for supernovae becomes less sensitive to all aspects of observing strategy, because not all systematics related to (for example) redshift and photometric classification have been folded in. We thus consider this ongoing research and instead introduce several proxy metrics that are expected to correlate with a full FoM including systematics.\\

{\bfseries Static Science Statistical FoM and $\langle N_i\rangle$:}
We developed a static science statistical FoM emulator based on a strategy's area and median depth, as shown in \autoref{fig:static_fom}. We also include the average number of $i$-band visits, $\langle N_i\rangle$, as a proxy for systematics in PSF modeling that lead to a preference for increased depth (or, at fixed depth, shorter exposure time).  There are three groupings: large area surveys which the FoM favours (on the right), deeper surveys which the systematics metric favours (near the top) and disfavoured surveys. We expect the statistical FoM to correlate with the final FoM which incorporates systematics, since most WL$+$LSS systematics are agnostic to survey strategy or prefer larger area (like the statistical FoM), unlike $\langle N_i\rangle$. Moving the 18000 deg$^2$ footprint allows us to maximize the statistical FoM without sacrificing $\langle N_i\rangle$; otherwise, we must weigh the relative importance of these two factors, which remains an area of active research. \\

{\bfseries Number and Redshift of Well-Measured Type Ia Supernovae}:
We impose a series of quality cuts on a simulated set of supernovae and find the redshift, $z_{\rm{cut}}$, above which a SN with the most likely stretch and color characteristics does not pass these cuts. The number of supernovae for which $z<z_{\rm{cut}}$ produces a sample of well-measured supernovae that will be excellent for cosmological constraints but will also ensure accurate classification, redshift estimation and reduce other systematic effects.\\

{\bfseries Cosmology with Rare Transients}:
We introduce two metrics for rare transient: the number of strongly lensed SNIa's and the number of kilonovae detected. In both cases, we assume LSST would act as a discovery machine, rather than requiring well-measured light curves, as it is feasible to individually follow up good candidates. These are combined with the supernova metric in \autoref{fig:transients}, which shows that regular visits are critical for both supernovae and kilonovae. 

\begin{minipage}{0.99\columnwidth}
    \centering
    \includegraphics[width=0.7\columnwidth,  trim={0cm 0cm 1.3cm 1.3cm}, clip]{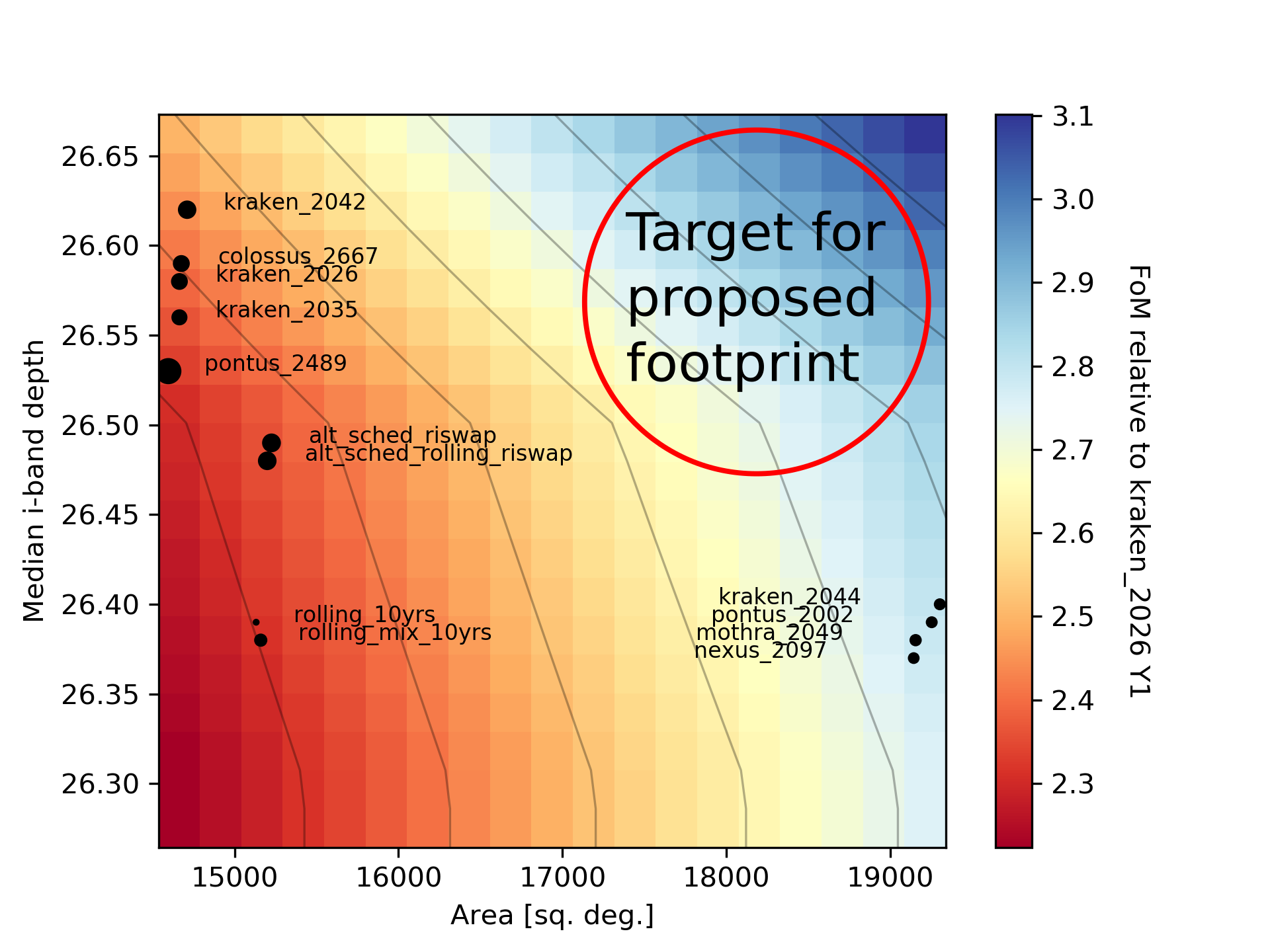}
    \captionof{figure}{Emulated statistical DETF FoM for the joint static science probes as a function of area and depth for Y10, relative to the FoM for \texttt{kraken\_2026} Y1. Lines represent constant FoM values and points are scaled by a weak lensing systematic metric where larger circles are preferable. Stage III priors are not included here while photo-$z$ errors are.
    }
    \label{fig:static_fom}
\end{minipage}\\

\vspace{10pt}
\begin{minipage}{0.99\columnwidth}
    \centering
    \includegraphics[width=0.93\columnwidth]{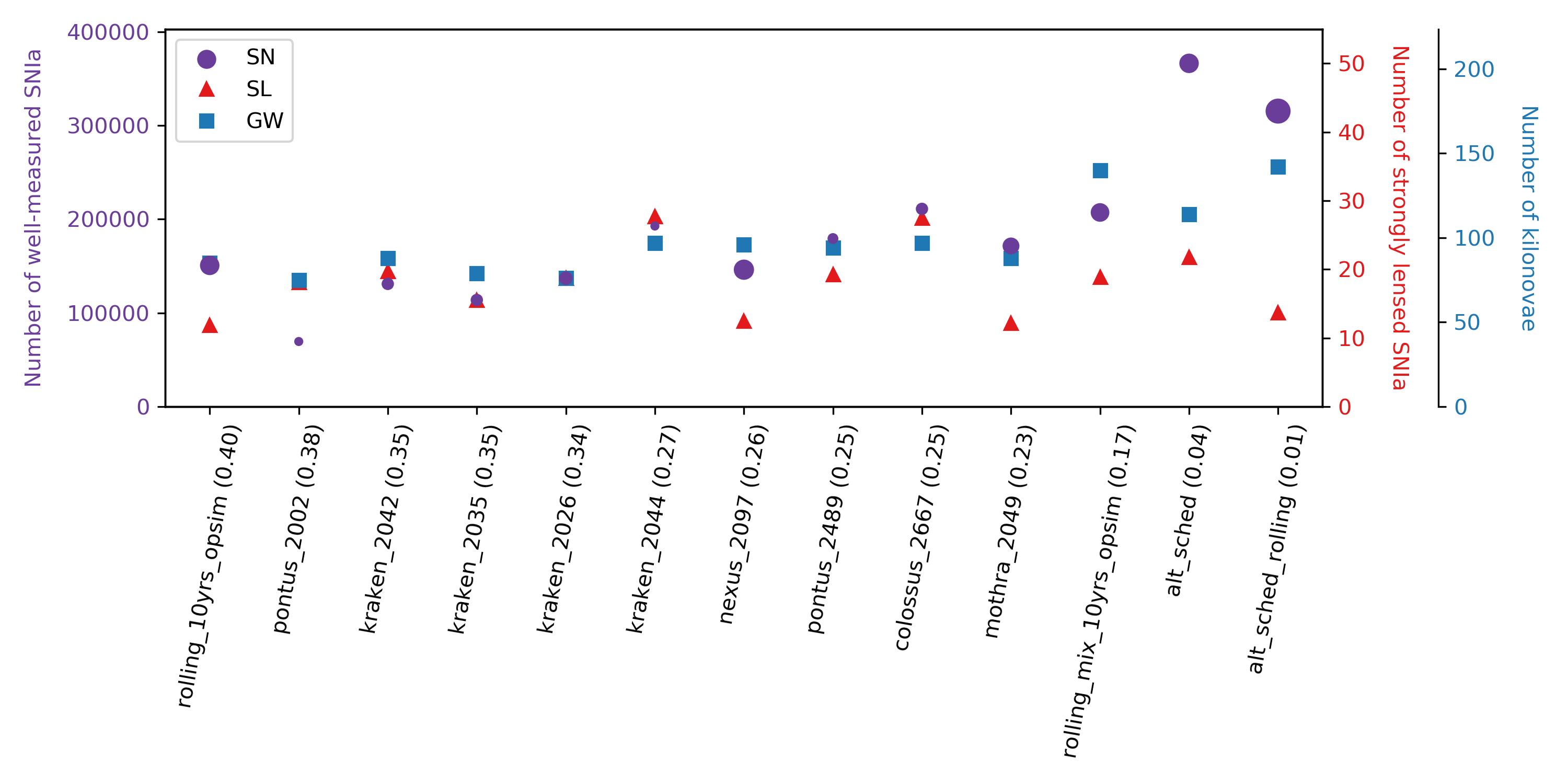}
    \captionof{figure}{Number of kilonovae (GW), strongly lensed type Ia supernovae with well-measured time delays (SL) (both assuming follow-up with other telescopes) and well-measured type Ia supernovae (SN) (scaled by $z_{\rm{cut}}$) for Y10 as a function of observing strategy, ordered by the percentage of visits in $r$-band separated by more than 15 days (in brackets).}  

    \label{fig:transients}
\end{minipage}

\vspace{.6in}

\section{Special Data Processing}

We would request the Data Management pipeline include a final co-add in all data releases of WFD with only a subset of DDF visits to produce a more homogeneous final survey. We also would be interested in best-seeing co-adds.

\section{Requests for New OpSim Runs}
\label{sec:requests}
In the course of our study, we determined several factors that may impact cosmology but which require further investigation. Here we describe several new simulations we would like to request. We request that all DDFs and mini-surveys be simulated together to study the impact on WFD. 
\begin{itemize}
    \item {\bfseries Seeing prioritization:} Static science will benefit from reserving the best seeing conditions for $r$ and $i$ band, but this may negatively impact strong lensing that tends to observe bluer objects. We thus request one simulation with $r$ and $i$ prioritized for best seeing and one with $g$, $r$ and $i$ prioritized to investigate this further.
    
    \item {\bfseries Variable exposure time:} We propose to allow exposure time to vary based on observing conditions (seeing, airmass, sky brightness and transparency) to \review{improve single-visit depth for transient detection and} achieve more uniform depth for galaxy detection/shape measurement. We request that pairs of visits (which we argue should be in different filters) be kept to the same exposure time and that all visits be longer than 15s but less than 60s, even when atmospheric conditions are such that the adaptive exposure time strategy would suggest otherwise. 
    
    \item {\bfseries Shorter exposure times:} To further investigate the effect of exposure time, we request an simulation similar to \texttt{pontus\_2489} except with 20s exposures in $gri$ while leaving 30s exposures in the other bands.
    
    \item {\bfseries Redistribution of filters:} We propose halving the number of visits in $y$-band and redistributing them. Figure 6 in \cite{graham2017} shows that while photo-$z$ are sensitive to changes in u-band depth, even halving the number of visits in $y$-band does not noticeably reduce photo-$z$ quality. This may be one of the reasons why AltSched performs so well for supernovae. \review{However this effect must be further explored before we can understand the full impact on photo-$z$.} Thus we would like to request simulations exploring the reduction of the number of visits in $y$-band by half and redistributing these to $griz$. The optimal filter distribution is unknown so multiple simulations would be appreciated.
    
    \item {\bfseries Clustered $u$ and  $y$  band visits:} Since $griz$ bands are \review{the most important} for our transient science cases, we request an observation which clusters (for example) all u-band and $y$-band visits in a month in a couple of nights (i.e., making all $u$-band in a short sequence of dark nights and making $y$-band visits a few weeks later around bright time), thus allowing $griz$ cadence to be improved in the intervening time. The effectiveness of this approach will depend on how important the different filters are for early classification which is still under investigation.
    
    \item{\bfseries Realistic seeing and weather:} In our analysis, it was found that a more realistic seeing model has a large impact on LSST science (Neilsen et al.~in prep), and we recommend updating the seeing and weather models (especially making some worst-case scenario weather simulations) in the next round of simulations. It would be interesting to consider strategies that are able to adapt in the event of poor weather.
    
    \item{\bfseries Rolling cadence:} We would like to work with the OpSim team to continue investigating rolling cadence as a promising avenue to improve transient cosmology. In particular, we would be interested in a rolling cadence that achieves uniformity in years 1, 3, 6, 10 and rolls in the intervening years, as well as rolling cadence options that maintain a low-level of uniform progression (e.g., 25\% of the baseline visits) in the deprioritized sky region.
    
    \item{\bfseries AltSched-like simulations:} The alternative scheduler to OpSim has shown impressive results for transient science due to its highly regular cadence in every filter (see \autoref{fig:transients}). We would thus request a simulation using OpSim, but with the same scheduling pattern as AltSched. This requires scanning the meridian, deviating only as necessary to increase season length. Observations should be taken in 45 minute blocks, incrementing the filter each revisit. The sky should be partitioned into two parts, observing each part on alternating nights. Ideally, the starting filter should be incremented by 2 each night (e.g., from $u$ to $r$ or $g$ to $i$). 
    
\end{itemize}
\section{Acknowledgements}
The DESC acknowledges ongoing support from the Institut National de Physique Nucl\'eaire et de Physique des Particules in France; the Science \& Technology Facilities Council in the United Kingdom; and the Department of Energy, the National Science Foundation, and the LSST Corporation in the United States.  DESC uses resources of the IN2P3 Computing Center (CC-IN2P3--Lyon/Villeurbanne - France) funded by the Centre National de la Recherche Scientifique; the National Energy Research Scientific Computing Center, a DOE Office of Science User Facility supported by the Office of Science of the U.S.\ Department of Energy under Contract No.\ DE-AC02-05CH11231; STFC DiRAC HPC Facilities, funded by UK BIS National E-infrastructure capital grants; and the UK particle physics grid, supported by the GridPP Collaboration.  This work was performed in part under DOE Contract DE-AC02-76SF00515.

This paper has undergone internal review in the LSST Dark Energy Science Collaboration. The internal reviewers were Jonathan Blazek, Chihway Chang, Ariel Goobar, William Hartley Isobel Hook, Saurabh Jha, Nacho Sevilla Noarbe, Anže Slosar, Eli Rykoff and Peter Yoachim. 
\section{References}
\bibliographystyle{hunsrt}
\begingroup
\renewcommand{\section}[2]{}%
\bibliography{refs}
\endgroup
\section*{Author Affiliations}
\parbox{\textwidth}{
\footnotesize
$^{1}$ African Institute for Mathematical Sciences, 6 Melrose Road, Muizenberg, 7945, South Africa\\
$^{2}$ South African Radio Astronomy Observatory, The Park, Park Road, Pinelands, Cape Town 7405, South Africa\\
$^{3}$ Kavli Institute for Cosmological Physics, University of Chicago, Chicago, IL 60637, USA\\
$^{4}$ Department of Physics and Astronomy, Rutgers University, Piscataway, NJ 08854, USA\\
$^{5}$ LPNHE, IN2P3-CNRS, 4 place Jussieu, 75005 Paris, France\\
$^{6}$ Laboratoire de Physique de Clermont (LPC) , CNRS/IN2P3, F-63000 Clermont-Ferrand, France\\
$^{7}$ McWilliams Center for Cosmology, Department of Physics, Carnegie Mellon University, Pittsburgh, PA 15213, USA\\
$^{8}$ Department of Physics, Stockholm University, AlbaNova, SE-10691 Stockholm, Sweden\\
$^{9}$ The Oskar Klein Centre for Cosmoparticle Physics, Department of Physics, Stockholm University, AlbaNova, 10691 Stockholm, Sweden\\
$^{10}$ Max-Planck-Institut f\"{u}r Astrophysik, Karl-Schwarzschild-Str.~1, 85748 Garching, Germany\\
$^{11}$ Physik-Department, Technische Universit\"{a}t M\"{u}nchen, James-Franck-Stra\ss{}e~1, 85748 Garching, Germany\\
$^{12}$ Department of Astronomy, University of Washington, Box 351580, U.W., Seattle, WA 98195, USA\\
$^{13}$ Department of Astronomy \& Astrophysics, Dunlap Institute for Astronomy \& Astrophysics, University of Toronto\\
$^{14}$ Dunlap Institute for Astronomy \& Astrophysics, University of Toroto\\
$^{15}$ University of Arizona\\
$^{16}$ University of California, Berkeley \\
$^{17}$ Mullard Space Science Laboratory, Department of Space and Climate Physics, University College London, Holmbury Hill Rd, Dorking RH5 6NT, UK\\
$^{18}$ SNSF Ambizione, Laboratory of Astrophysics, \'{E}cole Polytechnique F\'{e}d\'{e}rale de Lausanne (EPFL)\\
$^{19}$ Center for Cosmology and Astroparticle Physics, Ohio State University\\
$^{20}$ University of Chicago\\
$^{21}$ Institute of Cosmology and Gravitation, University of Portsmouth, Portsmouth, PO1 3FX, UK\\
$^{22}$ Physics Department, Lancaster University, LA1 4YB, U.K.\\
$^{23}$ University of Pennsylvania\\
$^{24}$ Lawrence Berkeley National Laboratory, Berkeley CA 94720 USA\\
$^{25}$ Kavli Institute for Particle Astrophysics and Cosmology, P.O. Box 20450, MS29, Stanford, CA 94309, USA\\
$^{26}$ Laboratoire de l'Acc\'{e}l\'{e}rateur Lin\'{e}aire, IN2P3-CNRS, Universit\'{e} de Paris-Sud, BP 34, 91898 Orsay Cedex, France\\
$^{27}$ University of Pittsburgh, Department of Physics and Astronomy and PITT PACC\\
$^{28}$ Department of Physics \& Astronomy, University College London, Gower Street, London WC1E 6BT, UK\\
$^{29}$ Centre for Astrophysics and Cosmology, University of Nova Gorica, Vipavska 11c, 5270 0 Ajdov\v{s}\u{c}ina, Slovenia \\
$^{30}$ NASA Jet Propulsion Laboratory, California Institite of Technology, Pasadena, CA 91109\\
$^{31}$ Centro de Investigaciones Energ\'{e}ticas, Medioambientales y Tecnol\'{o}gicas (CIEMAT), Av. Complutense 40, 28040, Madrid, Spain\\
$^{32}$ Physics Department, Brookhaven National Laboratory,  Upton, NY 11973\\
$^{33}$ Institute of Astronomy and Astrophysics, Academia Sinica, 11F of ASMAB, No.1, Section 4, Roosevelt Road, Taipei 10617, Taiwan\\
$^{34}$ Department of Physics, University of California, Davis\\
$^{35}$ University of Washington\\
}

\end{document}